\documentclass[aps,pra,showpacs,twocolumn]{revtex4}
\usepackage{graphicx}
\usepackage{epstopdf} 
\usepackage{amsmath}
\usepackage{amsfonts}

\begin{document}

\title{Dicke-type phase transition in a multimode optomechanical system}


\author{Jesse Mumford}
\affiliation{Department of Physics and Astronomy, McMaster University, 1280 Main St.\ W., Hamilton, ON, L8S 4M1, Canada}
\author{D. H. J.  O'Dell}
\affiliation{Department of Physics and Astronomy, McMaster University, 1280 Main St.\ W., Hamilton, ON, L8S 4M1, Canada}
\author{Jonas Larson}
\affiliation{Department of Physics,
Stockholm University, AlbaNova University Center, Se-106 91 Stockholm,
Sweden}
\affiliation{Institut f\"{u}r Theoretische Physik, Universit\"{a}t zu
  K\"{o}ln, K\"{o}ln, De-50937, Germany}

\date{\today}

\begin{abstract}
We consider the ``membrane in the middle'' optomechanical model consisting of a laser pumped cavity which is divided in two by a flexible membrane that is partially transmissive to light and subject to radiation pressure. Steady state solutions at the mean-field level reveal that there is a critical strength of the light-membrane coupling above which there is a symmetry breaking bifurcation where the membrane spontaneously acquires a displacement either to the left or the right. This bifurcation bears many of the signatures of a second order phase transition and we compare and contrast it with that found in the Dicke model. In particular, by studying limiting cases and deriving dynamical critical exponents using the fidelity susceptibility method, we argue that the two models share very similar critical behaviour. For example, the obtained critical exponents indicate that they fall within the same universality class. Away from the critical regime we identify, however, some discrepancies between the two models. Our results are discussed in terms of experimentally relevant parameters and we evaluate the prospects for realizing Dicke-type physics in these systems.  
\end{abstract}

\maketitle

\section{Introduction}\label{sec1}
Over the last decade optomechanics has emerged as a new playground where macroscopic
material oscillators coupled to light can display quantum phenomena.
In these hybrid systems, mechanical oscillators such as membranes or mobile mirrors
are coupled to light in optical resonators which allows them to be cooled to the quantum regime~\cite{optorev}. It has long been known that a light-matter coupling introduces a nonlinearity that can give rise to {\it optical bistability} characterized by drastic changes in the cavity field amplitude upon small parameter changes~\cite{meystre0}. Optical bistability has traditionally been discussed in cavity quantum electrodynamics in terms of a single optical resonator mode interacting on or near resonance with a thermal gas of two-level atoms~\cite{bis}. At the classical level this optical bistability, which displays a {\it hysteresis effect}, mimics a first order phase transition (PT). The advent of experiments with ultracold atomic gases trapped inside optical cavities has spurred a new generation of optomechanical investigations. In particular, in the dispersive regime  the atoms' \emph{external} degrees of freedom in the form of collective mechanical motion (phonons) are excited by the light and play a role closely analogous to the mechanical excitations of a mobile mirror \cite{bistabBerkley,beccav1}. The first round of cold atom experiments studied runaway instabilities such as superradiant Rayleigh scattering and collective atomic recoil lasing \cite{inouye,yoshikawa,slama}, however subsequent experiments have been able to realize the steady state phases of light and matter associated with these phenomena, and in particular, continuous transitions ({\it i.e.} second order PTs) have been observed~\cite{black03,beccav2,beccav3} as the atom-light coupling is varied. More precisely, a spontaneous {\it self-organization} \cite{domokos02,nagy08,nagy2010,piazza13} of the atomic cloud into a supersolid lattice structure  has been identified with the so called {\it Dicke PT}  (in fact, the transition observed in the experiments is most likely an ``open'' or classical version of the Dicke PT \cite{nagy11,bhaseen12} due to the fact that the optical cavity mirrors are not perfectly reflecting). The purpose of the original model studied by Dicke \cite{dicke} was to describe the collective emission of light by $N$ two-level atoms identically coupled to a single mode of the electromagnetic field. The change from the {\it normal} to the {\it superradiant} state as the atom-light coupling strength is varied was later identified as a continuous PT by Hepp and Lieb \cite{hepp}, and Wang and Hioe \cite{wang}.   In the dispersive cold atom experiments the role of the two internal atomic states is played by two momentum states of the atoms. The many similarities between the atomic optomechanical systems and the ones comprised of solid state mechanical oscillators suggests that related critical phenomena may also be found in the latter. This is the question we will address in the present work, and we will indeed demonstrate that continuous dynamical PT's similar to Dicke PT's might be realizable in state-of-the-art solid state optomechanical systems. 

\begin{figure}[t]
\includegraphics[width=8cm]{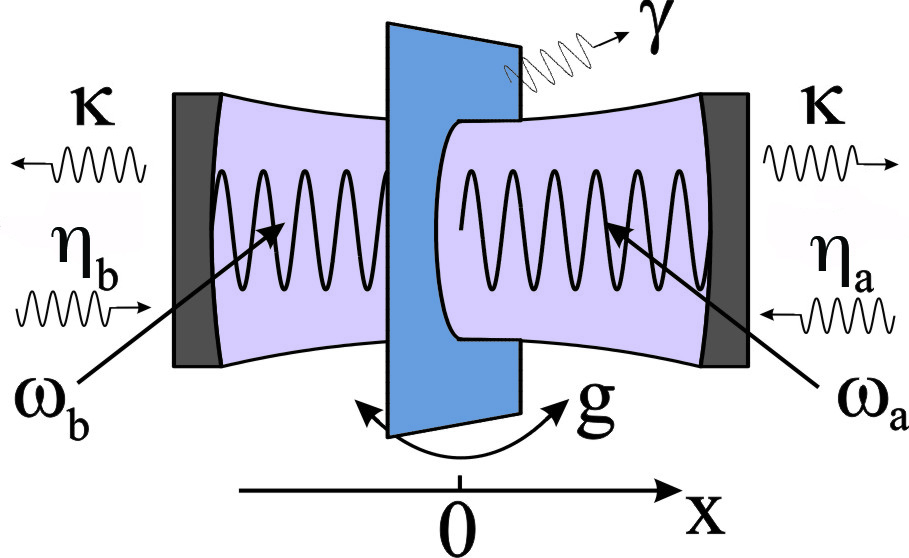}
\caption{A schematic depiction of the experimental setup realized in \cite{harris1} consisting of a partially transmissive flexible membrane inside a pumped Fabry-P\`erot cavity. The frequencies of the right and left cavities are $\omega_{a}$ and $\omega_{b}$, and they are separately pumped at rates $\eta_{a}$ and $\eta_{b}$, respectively. Photons are transmitted through the membrane at rate $g$, and the leakage of photons through each of the end mirrors is assumed to be equal and occur at rate $\kappa$. The direct mechanical decay rate $\gamma$ of the membrane will be discussed in Section \ref{sec:criticalexponents}. } \label{fig0}
\end{figure}

As a model system we choose the ``membrane in the middle'' setup realized in a pioneering set of experiments by the Yale group \cite{harris1,jayich08,sankey10} and studied theoretically in numerous papers, {\it e.g.}~\cite{meystre,heinrich,jonas1,duncan1,comptesrendus,marquardt13}. The membrane consists of a very thin film of dielectric material mounted on a frame and placed in the middle of a high-$Q$ optical Fabry-P\`erot cavity which is itself pumped by a laser, as depicted schematically in Fig.\ \ref{fig0}. The membrane is partially transmissive to light and is elastic like a drumhead so that when it is subjected to radiation pressure the membrane can be displaced to the left or right. However, the position of the membrane also determines the effective lengths of the left and right sub-cavities which means that there is a strong back-action in the system: the photon fields affect the state of the membrane, but changing the state of the membrane in turn alters the photon states. A similar back-action is found in the Dicke model (DM)~\cite{dicke,hepp,wang}, and indeed, we show in Section \ref{sec2} that by utilizing Schwinger's spin-boson mapping this model is identical to the Dicke one when losses and driving are excluded. Here, the Dicke PT manifests itself as spontaneous symmetry breaking of the position of the membrane, so that it is either displaced to the left or the right above a critical value of the light-membrane coupling strength. In section \ref{sec3} we derive an effective double well potential for the membrane, analogous to that which appears in the free energy in the Landau theory of continuous PT's, and use it to model the symmetry breaking. For the present system, the Dicke PT would imply that a macroscopic number of photons could be generated even in a closed undriven system. Reaching such regimes seems unlikely experimentally, and we therefore focus on the more realistic situation where the two photon modes are driven by two classical sources (lasers). Even though the resulting model is not identical to the Dicke Hamiltonian, we show in Section \ref{sec4} that on the mean-field level it can still be critical and that the critical exponents agree with those of the DM. The connection to the DM is further strengthened through study of the linearized collective excitations around the mean-field solution, which we perform in Section \ref{sec5}. We find that the basic structure of the spectrum as a function of the membrane-light coupling is the same in both models. This suggests that the Dicke type behaviour is not only manifested at the mean-field level, but also at a quantum level. By considering experimentally relevant parameters in Section \ref{sec6}, we thereby suggest that Dicke physics could be accessible in multimode optomechanical systems. Away from the universal critical regime there are, however, some differences between the two models. The most striking one is that in the limit of infinitely strong light-membrane coupling the state of the system returns to the normal one ({\it i.e.}  the non ``superradiant'' phase) with no membrane excitations present.  We conclude this paper in Section \ref{sec7} where we briefly discuss a no-go theorem for the PT in the DM and explain how it is circumvented here. We have also included an Appendix where we discuss Schr\"{o}dinger cat states of the membrane position.

\section{Model system}\label{sec2}
Within the two-mode approximation for the light, only one cavity mode to the right (frequency $\omega_{a}$) and one to the left (frequency $\omega_{b}$) of the membrane are taken into account. This is the simplest example of a multimode system. When the membrane is located in the centre of the cavity the modes have identical (bare) frequencies $\omega_{a}=\omega_{b}=\omega_{\mathrm{centre}}$. The two cavity modes are assumed to be
pumped at frequency $\omega_{\mathrm{centre}}$ by two lasers with respective real amplitudes $\eta_a$ and $\eta_b$, {\it i.e.}\ the cavity-pump detuning $\delta=0$ when the membrane is in the ground state. As an alternative, one could also pump through just one mirror at the same frequency $\omega_{\mathrm{centre}}$ since this is also the frequency of a ``global'' mode that extends throughout the cavity, but having two pumps would give more control. Photons are transmitted through the membrane between the $a$ and $b$ modes at a rate $g$ ($>0$). One-dimensional toy models that assume the optical properties of the membrane can be described by a $\delta$-function dielectric spike give $g$ as 
\begin{equation}
g \approx \frac{c}{L} \sqrt{\frac{1-R}{R}}
\end{equation} 
(see Eq.\ (14) in \cite{duncan1}), where $c$ is the speed of light, $L$ is the total length of the cavity and $R$ is the reflectivity of the membrane which is assumed to be close to unity.   

The membrane itself is assumed to behave as a harmonic oscillator of natural frequency $\omega$. It also experiences a radiation pressure proportional to the photon number difference between the two modes. 
Depending upon whether the membrane is positioned at a node or an anti-node of the bare cavity modes, the radiation pressure is, to a good approximation, either linear or quadratic in the displacement of the membrane~\cite{harris1,jayich08,meystre}. In this work only the linear coupling case will be considered. In a frame rotating at the pump/light mode frequency $\omega_{\mathrm{centre}}$ (and assuming the adiabatic approximation for the cavity modes~\cite{law}) the Hamiltonian is  
\begin{equation}\label{ham0}
\begin{array}{lll}
\hat{H} & = & \displaystyle{\frac{\hat{p}^2}{2m}+\frac{m \omega^2 \hat{x}^2}{2}+\hbar g\left(\hat{a}^\dagger\hat{b}+\hat{b}^\dagger\hat{a}\right)} \\  \\ & & +  \displaystyle{ \frac{2}{L} \hbar \omega_{\mathrm{centre}} \hat{x}\left(\hat{n}_a-\hat{n}_b\right) } \\ \\
& & \displaystyle{+\hbar \eta_a \left(\hat{a}+\hat{a}^\dagger\right)+\hbar \eta_b \left(\hat{b}+\hat{b}^\dagger\right),}
\end{array}
\end{equation}
see, {\it e.g.}, Eq.\ (1) in Reference \cite{comptesrendus}.
Here, $\hat{x}$ and $\hat{p}$ are the canonical position and momentum operators for
the membrane with $x=0$ being its expected position in the ground state in the absence of photons, and $\hat{a}$ and $\hat{b}$ ($\hat{a}^\dagger$ and
$\hat{b}^\dagger$) are the annihilation (creation) operators for the
two light modes obeying the bosonic commutation relations
$[\hat{a},\hat{a}^\dagger]=[\hat{b},\hat{b}^\dagger]=1$ and
$[\hat{a},\hat{b}]=0$. $\hat{n}_a=\hat{a}^\dagger\hat{a}$ and
$\hat{n}_b=\hat{b}^\dagger\hat{b}$ are the corresponding photon number
operators. The parameters in the Hamiltonian, plus the decay rates $\kappa$ and $\gamma$ accounting  for the cavity photon losses and membrane phonon losses, respectively, are
depicted schematically in Fig.\ \ref{fig0}.

PT's are properly defined in the thermodynamic limit
where the system becomes in some sense large. In order for our
Hamiltonian to maintain its form in this limit we apply the {\it Kac
prescription}~\cite{kac} which forces all terms to scale in the same way. To this end we introduce a dimensionless parameter $V$ which represents a characteristic measure of the size and put
\begin{equation}\label{ham}
\begin{array}{lll}
\hat{H} & = & \displaystyle{\frac{\hat{p}^2}{2}+\frac{\hat{x}^2}{2}+g\left(\hat{a}^\dagger\hat{b}+\hat{b}^\dagger\hat{a}\right)+\frac{\lambda}{\sqrt{V}}\hat{x}\left(\hat{n}_a-\hat{n}_b\right)}\\ \\
& & \displaystyle{+\eta_a\sqrt{V}\left(\hat{a}+\hat{a}^\dagger\right)+\eta_b\sqrt{V}\left(\hat{b}+\hat{b}^\dagger\right).}
\end{array}
\end{equation}
In this version of the Hamiltonian we have also made all quantities dimensionless, {\it i.e.}\ put $\hat{H} \rightarrow \hat{H}/\hbar \omega$, $\hat{x} \rightarrow \hat{x} \sqrt{\hbar/ m \omega}$, $\hat{p} \rightarrow  \hat{p} \sqrt{m \hbar \omega}$ and put all frequencies in units of the phonon frequency $\omega$ (and thereby also set the time scale). One interpretation of $V$ is as the number of photons in the system. In this way we see that $n_{a}$ and $n_{b}$ scale as $V$ and hence $\hat{a},\,\hat{b},\,\hat{x}, \, \hat{p} \sim \sqrt{V}$, meaning that all the terms in $\hat{H}$ are $\mathcal{O}(V)$.  For example, a coefficient such as $\eta \sqrt{V}$ can be interpreted as a scale dependent pumping $\eta'=\eta \sqrt{V}$, that is, as $V$ is increased the actual experimental pumping rate $\eta'$ should be increased in order that this term remain relevant. The scale independent rate $\eta$ is the pumping at some particular scale and is fixed. The parameter $\lambda$, which provides the coefficient of the radiation pressure term, is dimensionless and is given by
\begin{equation}
\lambda= \frac{2}{L}\frac{\omega_{\mathrm{centre}}}{\omega}
\sqrt{\frac{\hbar}{m \omega}} \ .
\label{eq:lambda}
\end{equation}

To make the parallel to the DM most transparent, let us begin by discussing the limiting case of $\eta_a=\eta_b=0$ where we regain the model Hamiltonian considered in numerous earlier works~\cite{harris1,jayich08,meystre,heinrich,jonas1,duncan1}. It directly follows that the total photon number operator $\hat{N}_{\mathrm{tot}}=\hat{n}_a+\hat{n}_b$ is preserved in this case which gives rise to a continuous $U(1)$ symmetry (corresponding to the operator $\hat{U}=\exp[-i \hat{N}_{\mathrm{tot}} \phi] $ which commutes with $\hat{H}$ for any $\phi$). To describe the situation where there is a net radiation pressure upon the membrane and it is displaced from $x=0$, {\it i.e.} in the symmetry broken phase, one can turn to a shifted basis by applying the displacement operator $\hat{D}(\alpha)=\exp(-i \alpha \hat{p})$ with $\alpha=\lambda(\hat{n}_a-\hat{n}_b)$ to the Hamiltonian. This leads to quartic terms in the photon operators which can be seen as ``photon-photon interaction terms'' generated by the collective coupling to the membrane~\cite{rabl}. As for the bosonic Josephson effect~\cite{becdw}, these effective interaction terms stabilize a {\it trapping state} where photon transmission through the membrane is suppressed as are the oscillations of the membrane itself~\cite{jonas1}. 


Another approach to the closed undriven model, seemingly overlooked in the past, is to use the fact that the number of photons is preserved and apply Schwinger's spin-boson mapping~\cite{sakurai} 
\begin{equation}
\begin{array}{lll}
\hat{S}_x=\hat{a}^\dagger\hat{b}+\hat{b}^\dagger\hat{a},\\ \\
\hat{S}_y=i\left(\hat{a}^\dagger\hat{b}-\hat{b}^\dagger\hat{a}\right),\\ \\
\hat{S}_z=\frac{1}{2}\left(\hat{n}_a-\hat{n}_b\right),
\end{array}
\end{equation}
giving the Hamiltonian in the spin representation
\begin{equation}\label{dicke}
\hat{H}_\mathrm{D}=\hat{n}_c+g\hat{S}_x+\sqrt{\frac{2}{V}}\lambda\left(\hat{c}+\hat{c}^\dagger\right)\hat{S}_z,
\end{equation}
where we have introduced the membrane creation (annihilation) operators $\hat{c}=\left(\hat{x}+i\hat{p}\right)/\sqrt{2}$ [$\hat{c}^\dagger=\left(\hat{x}-i\hat{p}\right)/\sqrt{2}$] and like before $\hat{n}_c=\hat{c}^\dagger\hat{c}$. Up to a trivial spin-rotation, $\hat{H}_\mathrm{D}$ is identical to the celebrated Dicke Hamiltonian~\cite{dicke} which at its heart describes $N$ spins (two-level atoms) coupled to a single harmonic oscillator (photon mode). The purpose of introducing the membrane operators $\hat{c}$ and $\hat{c}^\dagger$ is to make the analogue to the DM more evident~\cite{dickeqpt}. The Dicke PT~\cite{hepp,wang} from a  normal to a superradiant phase occurs at the critical coupling $\lambda_c=\sqrt{g}/2$ and corresponds to a breaking of spin symmetry and a macroscopic excitation of the harmonic oscillator. More precisely, it is the discrete $\mathbb{Z}_2$ symmetry characterized by 
\begin{equation}
(\hat{x},\,\hat{p},\,\hat{S}_x,\,\hat{S}_y,\,\hat{S}_z)\rightarrow(-\hat{x},\,-\hat{p},\,\hat{S}_x,\,-\hat{S}_y,\,-\hat{S}_z)
\end{equation}
that is broken.  The Dicke superradiant phase is therefore characterized by a non-zero $\langle\hat{x}\rangle$ and a non-zero $\langle\hat{S}_z\rangle$, \textit{i.e.} a breaking of symmetry to realize one of two options. The PT survives at zero temperature implying that the model is also quantum critical~\cite{dickeqpt}. In the DM the system ``size'' is identified with the total spin $\mathcal{S}$, and we see that all terms in the Dicke Hamiltonian scale as $\mathcal{S}$. 


It is worth appreciating the physical differences between the optomechanical system under discussion and the original formulation of the DM. In the latter, the magnitude $\mathcal{S}$ of the total spin is set by the number of two-level atoms. It is reasonable to assume that this is a certain fixed number. However, in our system the role of the atoms is played by photons, and in experimentally realistic situations photon losses are present.  These losses can be countered by pumping the cavity to give a fixed average photon number, but the steady state of a pumped, damped harmonic oscillator (the electromagnetic field) is in general a coherent state \cite{louisell} containing a superposition of different Fock states each corresponding to a different spin sector $\mathcal{S}$ in the DM. Furthermore, having a variable number of photons also makes it tricky to talk about a ground state. Despite these types of effects, it is believed that open systems can still be critical and display true PTs \cite{nagy2010,nagy11,bhaseen12,opencrit,goldbart09,sachdev}. The study of criticality in open systems is an interesting area in its own right, but exploring these foundational questions is not the main goal of the present work. Instead, we will show that novel features emerge when the two photon modes are pumped and a steady state is reached where pumping and losses are balanced. In Section \ref{sec5} of this paper we will estimate the effects of fluctuations arising from the open nature of the system.
 
We therefore turn to the the case of non-zero pump amplitudes $\eta_a$ and $\eta_b$. Adding the drive terms to the DM implies that the number operator $\hat{N}_{\mathrm{tot}}$ is no longer a constant of motion and we cannot utilize Schwinger's mapping in order to write the model as an effective spin Hamiltonian. Nevertheless, if $\eta_a=\pm\eta_b$ the $\mathbb{Z}_2$ symmetry of the DM still survives in the pumped system. In the boson representation the symmetry is characterized by
\begin{equation}\label{symm}
\left(\hat{x},\hat{p},\hat{a},\hat{b}\right)\rightarrow\left(-\hat{x},-\hat{p},\pm\hat{b},\pm\hat{a}\right),
\end{equation}
where the $\pm$ sign is determined by whether $\eta_a$ and $\eta_b$ have equal or opposite signs. Whenever $|\eta_a|\neq|\eta_b|$ this symmetry is, however, broken. Moreover, as will be demonstrated in Sec.~\ref{sec3} it follows from the fact that $g>0$ that this symmetry can only be spontaneously broken when $\eta_a=-\eta_b\equiv\eta$, \emph{i.e.} when there is a $\pi$ phase difference between the two pumps.

Having identified the $\mathbb{Z}_2$ parity symmetry of the driven Hamiltonian, it follows that the system may actually become critical in the thermodynamic limit at the point where the symmetry is spontaneously broken. However, as we are dealing with a driven system, we are careful to note that this is a {\it dynamical PT} appearing in the steady state. In the following we will additionally introduce losses which means that it is not really legitimate to talk about a proper quantum PT.

\section{Effective mean-field theory for the membrane}\label{sec3}
For experimentally relevant situations, the
cavity modes typically evolve on a faster time scale than the
mechanical membrane, {\it i.e.} $g \gg 1$ ~\cite{meystre,jonas1}.  (It should be noted that
in the figures and examples we give throughout this paper we have put $g \sim 1$ to better see the
effects of the symmetry breaking.  However, we expect there will be no
qualitative change in our results by having $g \gg 1$.) In this situation the optical modes can be adiabatically eliminated to give an effective model for the membrane alone. As we shall show below, the presence of the light induced forces leads to an effective double well potential for the membrane (when we are above the transition point). This approach makes it relatively easy to visualize the steady state properties of the system and it directly hints at the onset of a dynamical PT for certain light-membrane coupling strengths. In the following analysis we will also include cavity loss as a simple dissipative process (we neglect decoherence stemming from the losses). 

Using the above Hamiltonian~(\ref{ham}) we follow standard steps and derive the following Heisenberg equations of motion for the two optical boson operators 
\begin{eqnarray}
i \dot{\hat{a}} &=& \displaystyle{g \hat{b} + \frac{\lambda}{\sqrt{V}} \hat{x} \hat{a} + \eta_a\sqrt{V} -
i \kappa \hat{a}}, \nonumber \\ 
i \dot{\hat{b}} &=& g \displaystyle{\hat{a} - \frac{\lambda}{\sqrt{V}} \hat{x} \hat{b} + \eta_b\sqrt{V} -
i \kappa \hat{b}.}
\label{eq:eom}
\end{eqnarray} 
Here we have introduced a cavity decay rate $\kappa$ by assuming that photons can leak from the cavity into two independent zero temperature photon reservoirs. Since we assume a zero temperature bath and since we are mainly interested in the mean-field limit in this work we leave out reservoir induced fluctuations, the so called {\it Langevin noise terms}, in the above equations~\cite{gardiner}. At this stage we have not included phonon dissipation, but discuss this case briefly in the next section. The steady state solutions of the field operators, expressed in terms of the membrane displacement, become
\begin{eqnarray}
\hat{a}_\mathrm{ss} &=& - \frac{i \eta_a\sqrt{V} \kappa + g \eta_b\sqrt{V} + \hat{x} \eta_a
  \lambda}{g^2 + \kappa^2 + \hat{x}^2 \lambda^2/V} \nonumber \\ 
\hat{b}_\mathrm{ss} &=& - \frac{i \eta_b\sqrt{V} \kappa + g \eta_a\sqrt{V} - \hat{x} \eta_b
  \lambda}{g^2 + \kappa^2 + \hat{x}^2 \lambda^2/V} \, .
\label{eq:ss}
\end{eqnarray}
In this paper we only study steady state solutions to the mean-field equations of motion. However, single cavity optomechanical systems are known to also support self-sustained oscillations $x(t) \approx \bar{x}+A \cos(\omega t)$ about an average displacement $\bar{x}$ \cite{optomechanicalinstability}. Oscillating solutions also exist in the DM~\cite{bhaseen12}. They typically require a blue detuning of the pump laser from the cavity resonance whereas here we limit ourselves to the zero detuning case.

Introducing the effective potential for the membrane via 
\begin{equation}\label{effpot}
\dot{\hat{p}} = -\hat{x}-\frac{\lambda}{\sqrt{V}}\left(\hat{n}_a-\hat{n}_b\right)\equiv -\frac{d
V_{\mathrm{eff}}(\hat{x})}{d\hat{x}}
\end{equation}
and inserting the above steady state solutions for the field variables we find
\begin{eqnarray}
V_{\mathrm{eff}}(x) & = & \frac{x^2}{2} - \frac{2 g \eta_a \eta_bV}{g^2 +
  \kappa^2 + x^2 \lambda^2/V} + V\frac{(\eta_a^2 - \eta_b^2)}{(g^2 +
  \kappa^2)} \label{eq:Veff}   \\  \nonumber
&& \!\times\!\left [ \frac{\kappa^2 \mathrm{arctan}\!\left ( \frac{x
        \lambda/\sqrt{V}}{\sqrt{g^2 + \kappa^2}} \right )}{\sqrt{g^2 + \kappa^2}}\!
-\!\frac{g^2 x \lambda/\sqrt{V}}{g^2 \!+\! \kappa^2\! +\! x^2 \lambda^2/V}  \right ]\!. 
\end{eqnarray}
Since in the following we will consider the classical mean-field limit we drop the hat on the position operator $\hat{x}$. 

Equation (\ref{eq:Veff}) demonstrates how the $\mathbb{Z}_2$ parity symmetry is restored when $\eta_a=\pm\eta_b$ because then the antisymmetric terms in the square brackets do not appear and the remaining terms are symmetric in $x$.  As alluded to in the previous section, by choosing opposite signs $\eta_a=-\eta_b\equiv\eta$ only the lower lying photon state, which is antisymmetric for $g > 0$~\cite{duncan1}, is pumped when $x=0$. Indeed, making the transformation from the $a/b$ (right/left) representation to the $S/AS$ (symmetric/antisymmetric) representation, the pumping terms in Eq.\ (\ref{ham}) transform into
\begin{equation}
\begin{array}{lll}
\eta_a (\hat{a}^\dagger + \hat{a}) + \eta_b(\hat{b}^\dagger +  \hat{b})
 &\rightarrow& (\eta_a + \eta_b)(\hat{a}^\dagger_{S} + \hat{a}_S) \\ 
&&+ (\eta_a - \eta_b)(\hat{a}^\dagger_{AS} \!+\! \hat{a}_{AS}).
\end{array}
\end{equation}
Clearly, when $\eta_a=\eta_b$ the pumping of the antisymmetric mode vanishes. 

\begin{figure}[t]
\includegraphics[width=8cm]{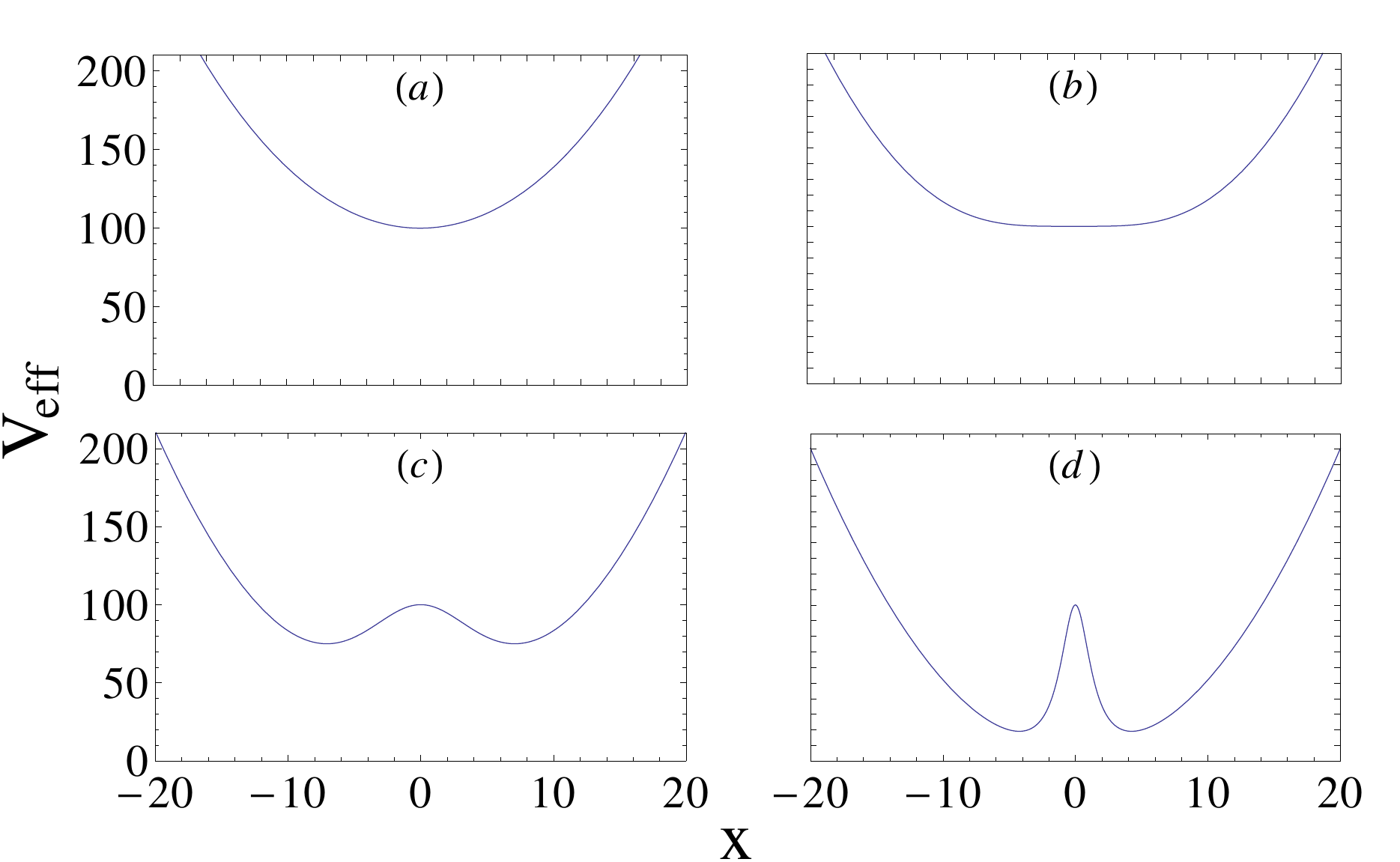}
\caption{The effective potential for the membrane as a function of the
membrane position for different values of $\lambda$; $\lambda = 0.5$
(a), $\lambda = 1$ (b), $\lambda = 2$ (c), and $\lambda = 10$ (d).
The other parameters are, $g = \kappa = \eta = 1$ (with $\eta_a =-
\eta_b$)  and $V =100$ giving $\lambda_c = 1$.  Note that (c) shows
the case when the splitting between the two minima is largest and in (d) the minima have
moved closer together again despite
increasing $\lambda$ further. As $\lambda\rightarrow\infty$ the central barrier becomes very narrow and the double well tends to a harmonic potential with a $\delta$-function at the centre.} \label{fig:Veff}
\end{figure}

In order to understand the physical significance of pumping the left and right modes with a $\pi$ phase difference we note that these are not stationary modes of the total cavity, \textit{i.e.}\ they are not stationary solutions of Maxwell's equations (even assuming perfect end mirrors),  but linear combinations of them are. Whereas the frequencies of the left and right modes cross at $x=0$, the global modes of the total cavity form an avoided crossing whose separation at $x=0$ is $2g$ \cite{harris1,jayich08,heinrich}. The gap $2g$ vanishes as the reflectivity of the membrane tends to unity. There is a close analogy between the double cavity problem for light and the double well problem for a quantum particle \cite{landau&lifshitz}, where the membrane displacement is equivalent to an imbalance between the depths of the two wells. However, there is also a crucial difference: whereas for massive particles tunneling through a potential barrier the lower energy eigenfunction of the double well is the symmetric combination of the left and right wave functions and the higher energy eigenfunction is the antisymmetric combination, for light passing through a dielectric slab the reverse is true---see Appendix A of \cite{duncan1} for a fuller explanation. Pumping with $\eta_{a}=-\eta_{b}$ excites the odd global mode which is the \emph{lower} frequency one.

In the following we shall disregard the case when $\eta_a=\eta_b$ because for $g >0$ the effective potential $V_{\mathrm{eff}}(x)$ can then only take the form of a single potential well centred at $x=0$ and the effect of increasing the light-membrane coupling $\lambda$ is just to narrow this single well. However,  when $\eta_a=-\eta_b$ we instead find that above a critical value
\begin{equation}
\lambda_c = \frac{g^2 + \kappa^2}{2 \eta \sqrt{g}}.
\label{eq:crit}
\end{equation}
the single well evolves into a double well, see Fig.\ \ref{fig:Veff}.  We thus obtain the classic Landau mean-field description of a continuous PT \cite{landau&lifshitzStat1}. By design, our scaling of the various terms in the Hamiltonian with system size results in $\lambda_{c}$ being independent of $V$. 

\begin{figure}[t]
\includegraphics[width=8cm]{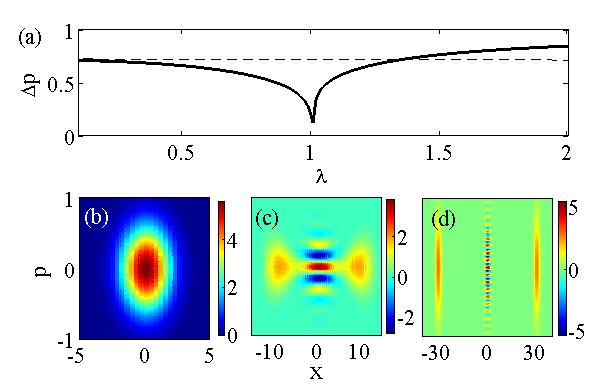}
\caption{Momentum squeezing (a) and the Wigner function for the membrane (b)-(d) for different light-membrane interaction strengths; $\lambda=0.95$ (b), $\lambda=1.005$ (c), and $\lambda=1.05$ (d). The rest of the parameters are, $g=\kappa=\eta=1$ (with $\eta_{a}=-\eta_{b}$) and $V=10000$. The critical point is $\lambda_c=1$ for these parameters, and the large value of $V$ means that the transition is clearly visible (the fact that $V$ is finite implies that maximum squeezing is not obtained for exactly $\lambda=1$). In (a) the dashed line gives the limit for classically accessible squeezing, {\it i.e.} that of a coherent state. The large squeezing in the vicinity of the critical coupling is clearly visible in the Wigner function shown in (c).} \label{fig:squeezing}
\end{figure}

According to quantum theory, the membrane can be in a superposition of being displaced to $+x_\mathrm{ss}$ or $-x_\mathrm{ss}$ which are the minima of the double well potential. Such superpositions are not allowed in the mean-field description which assumes a definite position for the membrane. Nevertheless, it is  instructive to make a brief excursion into the fully quantum world and examine what  properties such superpositions would have.  When the minima are separated by a macroscopic distance a {\it Schr\"{o}dinger cat state} is formed and the variance $\Delta x\equiv\sqrt{\langle x^2\rangle-\langle x\rangle^2}\gg1$. By Heisenberg's uncertainty principle we therefore expect that  the variance in the momentum quadrature obeys $\Delta p\equiv\sqrt{\langle p^2\rangle-\langle p\rangle^2}\ll1$. Indeed, in the DM it is known that the boson mode can become highly squeezed~\cite{nagy2010,nagy11,dickesq}. In Fig.~\ref{fig:squeezing} we plot some numerical results obtained by imaginary time-propagation of the Schr\"{o}dinger equation in the potential $V_\mathrm{eff}(x)$.  In particular, in Fig.~\ref{fig:squeezing} (a) we show how the momentum variance for the membrane depends upon the parameter $\lambda$. For $\Delta p<1/\sqrt{2}$ (marked by the dashed line) we achieve squeezing beyond the classical limit~\cite{mandel}, with $\Delta p$ reaching its minimum at the  critical point at $\lambda=1$.   In  Fig.~\ref{fig:squeezing} (b)-(d) we plot the Wigner phase space distribution $W(x,p)=\frac{1}{\pi}\int\,dy\psi^*(x+y)\psi(x-y)e^{2ipy}$ where $\psi(x)$ is the ``wave function'' of the membrane. The presence of a coherent superposition is signalled by the interference pattern at $x=0$ whose onset coincides with the formation of the double well.

Macroscopic superposition states and momentum squeezing of the membrane are of considerable fundamental interest but unfortunately  they would be hard to observe in practice. In the first place this is because we are dealing with an open system in which light (that is entangled with the state of the membrane) is leaking out of the cavity where is it rapidly measured by the environment.  This is a double-edged sword because it both allows the continuous monitoring of the state of the membrane and at the same time collapses the superposition. An analogous situation occurs in atomic gas experiments where the two symmetry broken states correspond to the occupation of two different sublattices inside the cavity and are distinguished by the cavity light having either a $0$ or $\pi$ phase difference with the pump light, something which can be seen using heterodyne detection \cite{beccav3}. A second consideration, which is of a more technical nature, is the degree to which we can protect the double well from explicit symmetry breaking terms in the Hamiltonian. In Appendix \ref{sec:appendix} we calculate at what asymmetry the cat state collapses into one well or the other due \textit{e.g.}, to slightly imbalanced pumping,  and find that this imposes severe experimental constraints.

\section{Dynamical criticality}\label{sec4}
\subsection{Emergence of criticality} 
\begin{figure}[t]
\includegraphics[width=8cm]{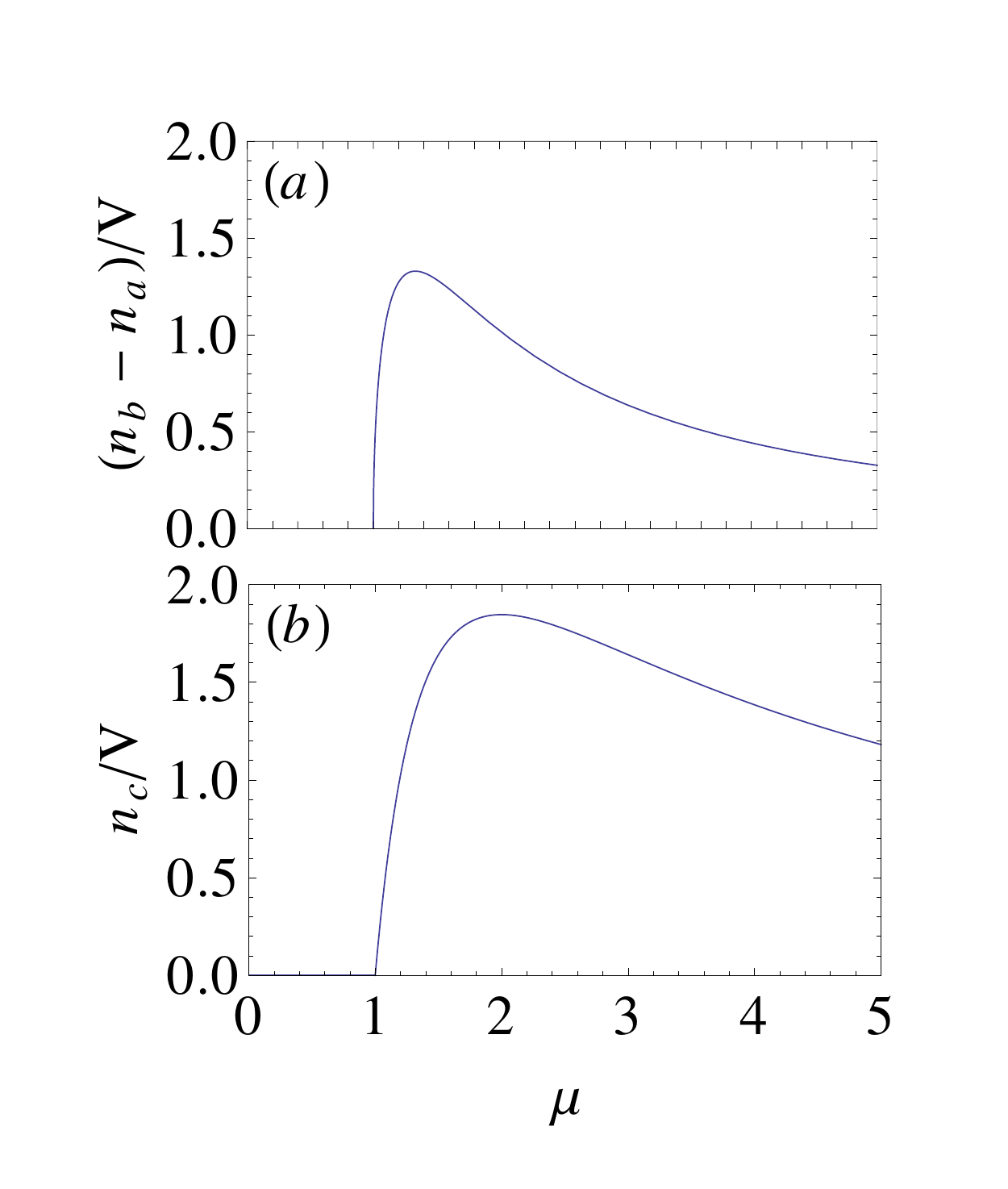}
\caption{Cavity photon number difference (a), and the number of
  phonons in the oscillator (b), both as a function of $\mu =
  \lambda/\lambda_{c}$.  The parameter values are $g = 3$, $\kappa =
  2$ and $\eta = 4$.  The symmetry broken phase is marked by
  macroscopic excitations in the phonon mode and a shift in the number
difference of photons between the two cavities.} \label{fig:amplitudes}
\end{figure}

In terms of the ratio $\mu = \lambda/\lambda_c$ of the light-membrane coupling to its critical value, the minima of $V_\mathrm{eff}(x)$ are at the membrane positions
\begin{equation}
x_{\mathrm{ss}\pm} = \begin{cases}
0, & \,   \quad \mu \leq 1 \\
\pm \sqrt{2 \epsilon_0} \frac{\sqrt{\mu - 1}}{\mu}, & \, \quad \mu
> 1,
\end{cases}
\label{eq:x}
\end{equation}
where  $\epsilon_0 = \frac{2 g \eta^2V}{g^2+\kappa^2}$. These are the fixed points (steady states) for $x$ of the mean-field Heisenberg equations of motion. Plugging Eq.\ (\ref{eq:x}) into Eq.\ (\ref{eq:Veff}) (with $\eta_a = -\eta_b$) we get the shift in the ``ground state'' energy 
\begin{equation}
\frac{E_0}{\epsilon_0} = \begin{cases}
 1, & \,  \mu \leq 1 \\
\frac{2 \mu - 1 }{\mu^2}, & \, \mu
> 1 \, .
\end{cases}
\label{eq:GSEnergy}
\end{equation}
In the thermodynamic limit, $V\rightarrow\infty$, the steady state
position (\ref{eq:x}) diverges, {\it i.e.}\ the membrane becomes
highly excited. This marks the presence of a continuous
PT with a critical coupling $\lambda_c$; in the symmetry
broken phase, $\mu>1$, the $\mathbb{Z}_2$ parity symmetry is broken
and the membrane is shifted to $|x_{\mathrm{ss}\pm}|$ in either
direction. The phase with $\mu<1$, where the membrane is in the
ground state, will be called {\it normal phase} in analogy with the
corresponding phase of the DM. Returning to Eq.~(\ref{eq:ss})
we see how the photon amplitudes $n_a=a_\mathrm{ss}^*a_\mathrm{ss}$
and $n_b=b_\mathrm{ss}^*b_\mathrm{ss}$ are altered across the
transition. In the normal phase the field amplitudes are given by the
standard result for a resonantly pumped dissipative oscillator, {\it
  i.e.}\ $n_a=n_b=\eta^2V/\left(g^2+\kappa^2\right)$. As expected, in
this phase the photon amplitudes are independent of the light-membrane
coupling $\lambda$ since $x_\mathrm{ss}=0$ and the light field has a
node at this point. In the symmetry broken phase the membrane is
shifted from $x=0$ which causes the pump to be effectively detuned from the dressed cavity
modes, and thereby the value of $n_a - n_b$ will be shifted in this
phase as well. The shifts in $n_a - n_b$ and $n_c$ (phonon number) can be seen in
Fig.\ \ref{fig:amplitudes} (a) and (b), respectively.  Note that $n_a\neq n_b$ due to the different numerators
in Eq.~(\ref{eq:ss}).

Let us now compare the present pumped and damped system to that of
the DM. Indeed, there are some differences between these
results and those obtained within the DM. In the next subsection we will
discuss this further where we also compare some critical
exponents of the present model to those of the DM. Instead,
here we compare non-universal properties away from the critical
regime. We first note that in the DM $\lambda_{c} \propto
g^{1/2}$~\cite{emary} while here we have $\lambda_{c} \propto
g^{3/2}$. Also, most strikingly, for $\lambda \gg\lambda_c$ the
quantities in Eqs.\ (\ref{eq:x}) and (\ref{eq:GSEnergy}) approach
zero, whereas in the DM they show a linear and quadratic
behaviour, respectively, and tend to infinity.  The asymptotic
behaviour can be seen in Fig.\ \ref{fig:amplitudes} where amplitudes
reach a maximum and decrease as we increase $\mu$.  As we saw in the
previous section, the imbalance $n_a-n_b$ serves as the $z$ component
of the large spin in the DM (when we set the pumping to
zero). From Eq.~(\ref{effpot}) we see that in steady state
$x_\mathrm{ss}\propto(n_a-n_b)$ implying that the displacement scales
in the same way as the photon imbalance and contrary to the Dicke
model the imbalance goes to zero for large light-membrane coupling
strengths. A physical explanation for this asymptotic behaviour is
that as the membrane shifts away from $x=0$ it changes the two cavity
lengths and moves them out of resonance with the pumping. Since the
pumping amplitude $\eta$ is held fixed as we vary $\lambda$ it means
that for very large light-membrane couplings very few photons at all
will be scattered into the cavity and the light pressure cannot
overcome the membrane's own spring constant. In this way the
membrane's displacement is reduced at large $\lambda$. The turning
point at which the reduction begins can be seen in Fig.\
\ref{fig:amplitudes}(b) where the membrane reaches maximal excitations
at $\mu = 2$.  Note that for the DM we do not encounter a similar situation since there is no driving mechanism present in it. We should also note that this transition in the large coupling limit is not a PT but a crossover.

Even though we are mainly concerned with the resonant driving case in this work, we just briefly state that for a small cavity pump detuning $\Delta=\omega_a-\omega_\mathrm{center}=\omega_b-\omega_\mathrm{center}$ the qualitative behaviour is not changed. In this case the off-resonant driving has the effect of shifting the critical coupling to
\begin{equation}
\lambda_{\Delta,c} = \frac{1}{2 \eta} \sqrt{\frac{\left[(g-\Delta)^2 +
  \kappa^2\right]\left[(g+\Delta)^2 + \kappa^2\right]}{g+\Delta}}.
\end{equation} 
The denominator of this expression shows the difference between red and blue pumping, and especially that the critical behaviour may be lost when $\Delta<-g$.

\subsection{Critical exponents}
\label{sec:criticalexponents}
\begin{figure}
\begin{center}
\includegraphics[width=8cm]{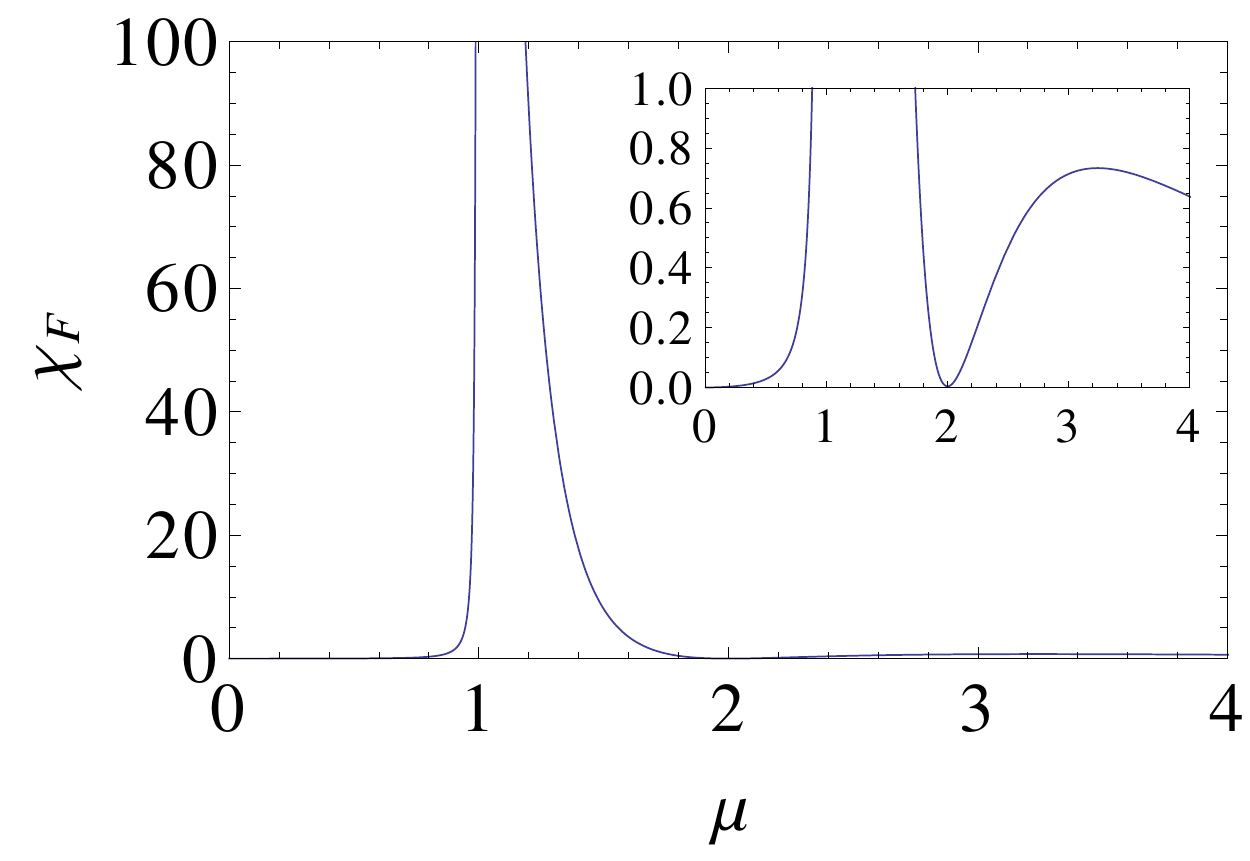}
\end{center}
\caption{Fidelity susceptibility $\chi_{F}$ as a function of $\mu=
  \lambda/\lambda_{c}$ for $\epsilon_0 = 100$. The inset shows the magnified region around $\mu
= 2$. As explained in the text, the vanishing susceptibility is an outcome of an interplay between the pumping and the light-membrane interaction energy. Even though the critical exponents for the susceptibility of our model are identical to those of the DM, this extra phenomenon away from criticality is not present in the DM.}
\label{fig:FSmu}
\end{figure}

We shall now attempt to identify some critical exponents and use them to compare the present dynamical PT to that of the DM. To do this we first consider the {\it fidelity susceptibility} (FS)~\cite{fid1,yang} which is a measure of how sensitive the system is to changes in some control parameter.
The FS is very similar to the well known magnetic susceptibility or heat capacity when the control parameters are the magnetic field or the temperature,
respectively. In particular, we expect a diverging response of the system to small parameter changes in the vicinity of the critical point $\lambda_c$. The FS is generically defined as~\cite{cozzini,buonsante}
\begin{equation}
\chi_{\mathrm{F}}(\lambda) = - \frac{1}{2} \frac{d^2}{d \delta^2} \langle
\psi_0(\lambda) \vert \psi_0(\lambda + \delta) \rangle
\mid_{\delta =0}
\label{eq:FS1}
\end{equation} 
where $\delta \ll 1$ in scaled units and $\psi_0$ is the ground state
wave function of the system. We shall apply the FS concept to the membrane wave function with the
light fields adiabatically eliminated, i.e.\ the membrane experiences the
effective potential given in Eq.\
(\ref{eq:Veff}). Furthermore, we approximate the membrane's wave function by a
Gaussian centred at $x = 0$ in the normal phase $\lambda
< \lambda_c$ and a Gaussian centred at one of the two nonzero
values in Eq.\ (\ref{eq:x}) in the symmetry broken phase, $\lambda >
\lambda_c$. We are thus quantizing the system. The Gaussian
approximation is valid as long as we stay away from the critical
region where the steady state wave function is in the process of
splitting.  When the double well is clearly formed we are out of this
critical region which happens when the distance between the two
minima, $x_{\mathrm{ss}+} - x_{\mathrm{ss}-}$ (see Eq.\ (\ref{eq:x})), is much greater than the width of the Gaussian.  To calculate the variance of the wave function we expand Eq.\ (\ref{eq:Veff}) around the steady state $x_{\mathrm{ss}}$ and identify the width $\sigma^2 = \frac{1}{2\sqrt{A}}$, 
where $A$ is the coefficient in front of the quadratic term in the expansion.  Explicitly, this gives the variance 
\begin{equation}
\sigma^2 = \begin{cases}
\frac{1}{2 \sqrt{1-\mu^2}}, & \,  \mu < 1 \\
\frac{1}{2 \sqrt{4 \left ( \frac{\mu - 1}{\mu} \right )}}, & \,  \mu
> 1
\end{cases}
\label{eq:sigma}
\end{equation}
which combined with Eq.\ (\ref{eq:FS1}) gives us the leading order terms in the FS as
\begin{equation}
\chi_{\mathrm{F}} = \begin{cases}
\frac{\mu^2}{16 (\mu^2 -1)^2}, & \,  \mu < 1 \\
\epsilon_0 \frac{\sqrt{\mu} (\mu - 2)^2}{8 \mu^5 \sqrt{\mu - 1}} +
\frac{1}{64 \mu^2 (\mu - 1)^2}, & \,  \mu
> 1.
\end{cases}
\label{eq:FS2}
\end{equation}
Below the critical coupling the position of the wave function is centred at $x=0$, so any changes in the FS derive entirely from how its width
is altered as $\lambda$ is varied. However, above the critical
coupling $\lambda$ affects both the size and the position of the wave
function. It follows that the leading order term is the
position-dependent term due to the factor of $V$ in $\epsilon_0$ which
is taken as large in the thermodynamic limit. From the analytical
results given in Eq.\ (\ref{eq:FS2}) we can extract the corresponding
critical exponents $\alpha_\pm$, where $\alpha_{+}$ is the exponent above the transition and $\alpha_{-}$ is the exponent below the transitions; for $\mu < 1$, $\chi_{\mathrm{F}}
\propto 1/(\mu-1)^2$ giving the exponent $\alpha_- = 2$, while for
$\mu >1$, $\chi_{\mathrm{F}} \propto 1/\sqrt{\mu - 1}$ the exponent
$\alpha_+ = 1/2$. These exponents are identical to those of the closed
DM \cite{liu}. In fact, they do not depend on $\kappa$ and so
are the same for the open and closed cases. The FS shows us that at the turning
point at $\mu = 2$ the system is rather insensitive to changes in $\lambda$; the system's response can only derive from higher order terms in the expansion. This non-universal feature of the behaviour of the susceptibility is illustrated in Fig.\ \ref{fig:FSmu}.

It is known that the steady state boson excitation (for the membrane) scales as $n_c\sim|\mu-1|^{-\beta}$ with $\beta=-1/2$ for the closed and $\beta=-1$ for the open DMs~\cite{nagy2010,nagy11,dickecrit0}. This discrepancy in scaling behaviour between the open and closed DMs has also been explored experimentally~\cite{dickecrit2} and it was found that the measured exponent is in closer agreement with the value $\beta=-1$. In the present system $\hat{c}_\mathrm{ss}=\lambda\left(n_a-n_b\right)/\sqrt{2V}$. Thus, for the steady state in the symmetry broken phase we have
\begin{equation}\label{phoncrit}
n_c\propto\left(n_a-n_b\right)^2\propto x_{\mathrm{ss}}^2\sim|\mu-1|,
\end{equation}
where the result given in Eq.~(\ref{eq:x}) has been used and also the fact that $x_\mathrm{ss}\propto(n_a-n_b)$. Equation (\ref{phoncrit}) tells us that our mean-field dynamical critical exponent $\beta=-1$ which is in perfect agreement with the result of the open DM. The fact that the exponents found for the fidelity susceptibility and the phonon number $n_c$ in the present model agree with those of the DM suggests, but is not a strict proof, that the two models belong to the same universality class. To fully characterize the universality class one would need to determine all independent exponents~\cite{goldenfeld} which is, however, beyond the scope of the present work.        

It is useful at this point to catalogue how the main dissipative processes appear in our mean-field treatment and the effects, if any, we expect them to have upon the critical behaviour. Firstly, the leaking of light from the cavity is a spontaneous process that can lead to membrane cooling \cite{optorev}, but since this decay channel acts on the light rather than directly upon the membrane we were able to include the average rate $\kappa$ as a parameter in our effective membrane potential in Eq.\ (\ref{eq:Veff}) where, along with the pumping rate, it simply accounts for the amount of light in the cavity. The adiabatic elimination of the light field dynamics needed to obtain Eq.\ (\ref{eq:Veff}) precludes the existence of frictional effects associated with membrane cooling because these require a time lag between the mirror and the light dynamics \cite{marquardt08,nagy09} (in the next section we shall look at small fluctuations around the adiabatic solution and see some evidence of cooling-type phenomena due to the existence of imaginary parts of the frequencies of these fluctuations). As we have seen above, by working with the FS for the membrane in an effective \emph{conservative} potential the leaking of light affects $\beta$ but not $\alpha$. Secondly, the membrane can also exchange phonons with the mount to which it is attached or couple to other modes of its motion other than the ones we want and this can lead to either direct heating or cooling of the membrane depending upon the effective relative temperature. We have not included these processes in our effective potential due to the fact that they act directly upon the membrane and so would lead to a non-Hermitian model. Even though the phonon reservoir is likely to contain thermal phonons at the relevant frequencies, here we will assume the simplified picture of a zero temperature reservoir and neglect fluctuations responsible for decoherence. Thus, as for the cavity fields we follow a master equation approach with the Langevin forces neglected. This might not give a quantitative description~\cite{bath}, but we will argue that it gives the correct qualitative picture. For a zero temperature phonon bath, the Heisenberg equations of motion for the membrane position $x$ and momentum $p$ will include damping terms of the form $-\gamma x/2$ and $-\gamma p/2$, respectively, where $\gamma$ is the membrane decay rate. Effectively this amounts to a complex membrane vibrational frequency, which will not break the $\mathbb{Z}_2$ symmetry. Thus, the equations of motion (\ref{eq:eom}) are still symmetric even with phonon damping taken into account. This observation is crucial for the PT to survive phonon losses. We have explored the effects of the terms $-\gamma x/2$ and $-\gamma p/2$ numerically by looking at the steady state solutions of the full set of semi-classical Heisenberg equations of motion. From this analysis we verified that the exponent $\beta$ remains the same even for a non-zero $\gamma$. The effect of $\gamma$ is mainly a rescaling of the critical coupling $\lambda_c$; the coupling is multiplied by the factor $\sqrt{1+\gamma^2}$. For zero $\eta_a$ and $\eta_b$, this prefactor can be demonstrated analytically~\cite{jonas1}. Physically this means that the transition will appear at larger couplings which is reasonable since the action of the membrane damping/friction has to be countered. Now, we may ask whether thermal phonons in the bath could alter the parity symmetry. However, at this level of mean-field approximation the inclusion of a ``thermal Lindblad term'' in the master equation will not change the equations of motion and hence the $\mathbb{Z}_2$ symmetry is preserved also in this case. Again, this is not surprising since a thermal state is Gaussian and does not favour a breaking of the parity in $x$ and $p$. This said, even if the dynamical PT is still present, once a phonon bath has been taken into account we cannot rule out the possibility that the universal properties of the transition could be different if phonon quantum fluctuations were to be included.

\section{Stability Analysis}\label{sec5}
In order to determine the stability of the mean-field (classical) solutions we follow the standard approach of linearizing the equations of motion around the steady state solutions. This will give us the stability and also the low-energy collective excitations around the mean-field solutions. Thus we expand the operators as
\begin{eqnarray}
\hat{a} &\rightarrow& \alpha_\mathrm{ss} + \delta \hat{a}, \nonumber \\
\hat{b} &\rightarrow& \beta_\mathrm{ss} + \delta \hat{b}, \nonumber \\
\hat{c} &\rightarrow& \gamma_\mathrm{ss} + \delta \hat{c},
\end{eqnarray}
from which we get the linear coupled equations for the fluctuations $\delta\hat{a}$, $\delta\hat{b}$, and $\delta\hat{c}$;
\begin{eqnarray}
\delta \dot{\hat{a}} &=& -i \frac{\lambda}{\sqrt{2V}} \alpha_\mathrm{ss}
(\delta \hat{c}^\dagger + \delta \hat{c}) - \left ( i \sqrt{\frac{2}{V}} \lambda
  \mathrm{Re} \left [\gamma_\mathrm{ss} \right ] + \kappa \right )
\delta \hat{a}  \nonumber \\ && - ig
\delta \hat{b}, \nonumber \\
\delta \dot{\hat{b}} &=& i \frac{\lambda}{\sqrt{2V}} \beta_\mathrm{ss}
(\delta \hat{c}^\dagger + \delta \hat{c}) - \left (i\sqrt{\frac{2}{V}} \lambda
  \mathrm{Re} \left [\gamma_\mathrm{ss} \right ]+ \kappa \right )
\delta \hat{b} \nonumber \\ && - ig
\delta \hat{a}, \nonumber \\
\delta \dot{\hat{c}} &=& -i \delta \hat{c} 
- i
\frac{\lambda}{\sqrt{2V}} \left [ \alpha_\mathrm{ss}\delta \hat{a}^\dagger+\alpha_\mathrm{ss}^*\delta \hat{a} -\beta_\mathrm{ss}\delta \hat{b}^\dagger-\beta_\mathrm{ss}^*\delta
  \hat{b} \right ]. \nonumber \\
\label{eq:lineom}
\end{eqnarray}
Here we have excluded trivial constants, and the steady state solutions
are given by $\alpha_\mathrm{ss}$, $\beta_\mathrm{ss}$, and
$\gamma_\mathrm{ss}$ (which should not be confused with the membrane
decay rate, $\gamma$). We shall restrict ourselves to the steady state solutions corresponding to the membrane at $x=0$.   Rather than working with the creation/annihilation operators, we study the quadratures,
$\delta \hat{X}_a \equiv (\delta \hat{a}^\dagger + \delta \hat{a})/\sqrt{2}$
and $\delta \hat{P}_a \equiv (\delta \hat{a}^\dagger - \delta{a})/i
\sqrt{2}$ and equivalently for the other two modes.  By introducing $u(t) = (\delta \hat{x}, \delta \hat{p}, \delta
\hat{X}_a, \delta \hat{P}_a, \delta \hat{X}_b, \delta
\hat{P}_b)^{\mathrm{T}}$ and substituting the steady state solutions of
$\alpha_\mathrm{ss} = -\beta_\mathrm{ss} = \eta\sqrt{V}/(g+i\kappa)$ and
$\gamma_\mathrm{ss} = 0$ into Eqs.\ (\ref{eq:lineom}) the linearized
equations can be written in the matrix form
\begin{equation}
\frac{d}{dt} u(t) = D u(t),
\end{equation}
where $D$ is the {\it drift matrix}
\begin{equation}
D = \begin{pmatrix}
 0 & -1 & 0 & 0 & 0 & 0  \\ 
 1 & 0 & \mu \sqrt{\frac{g}{2}} &\mu \frac{\kappa}{\sqrt{2g}} & \mu \sqrt{\frac{g}{2}} &
 \mu \frac{\kappa}{\sqrt{2g}}  \\ 
 -\mu \frac{\kappa}{\sqrt{2g}}& 0 & -\kappa & 0 & 0 & -g \\ 
 \mu \sqrt{\frac{g}{2}}& 0 & 0 & -\kappa & g & 0 \\ 
 -\mu \frac{\kappa}{\sqrt{2g}}& 0 & 0 & -g & -\kappa & 0 \\ 
 \mu \sqrt{\frac{g}{2}}& 0 & g & 0 & 0 & -\kappa 
\end{pmatrix}.
\label{eq:dmatrix}
\end{equation}
The eigenvalues of $iD$ will give the small excitation frequencies of
the system. The general expression is too cumbersome  to be given here and instead we focus on the limiting case of zero photon losses. For
$\kappa = 0$ we get the pairs of excitation frequencies
\begin{eqnarray}
\omega_{\mathrm{ex}} & = &\left(\pm g,\, \pm\frac{\sqrt{1+g^2+\sqrt{(g^2-1)^2+4g^2
      \mu^2}}}{\sqrt{2}},\right. \nonumber \\
& & \left.\pm\frac{\sqrt{1+g^2-\sqrt{(g^2-1)^2+4g^2
      \mu^2}}}{\sqrt{2}}\right).
\label{eq:freqs}
\end{eqnarray}
We see that the third pair becomes imaginary at $\mu = 1$ which
signals a classical pitchfork bifurcation.  Specifically one becomes positive and imaginary which
means small perturbations will cause exponential growth away from the
initial position.  This motion defines the process of the system
moving from the newly unstable position at the centre of the double
well to one of the two minima. 

\begin{figure}[t]
\begin{center}
\includegraphics[width=8cm]{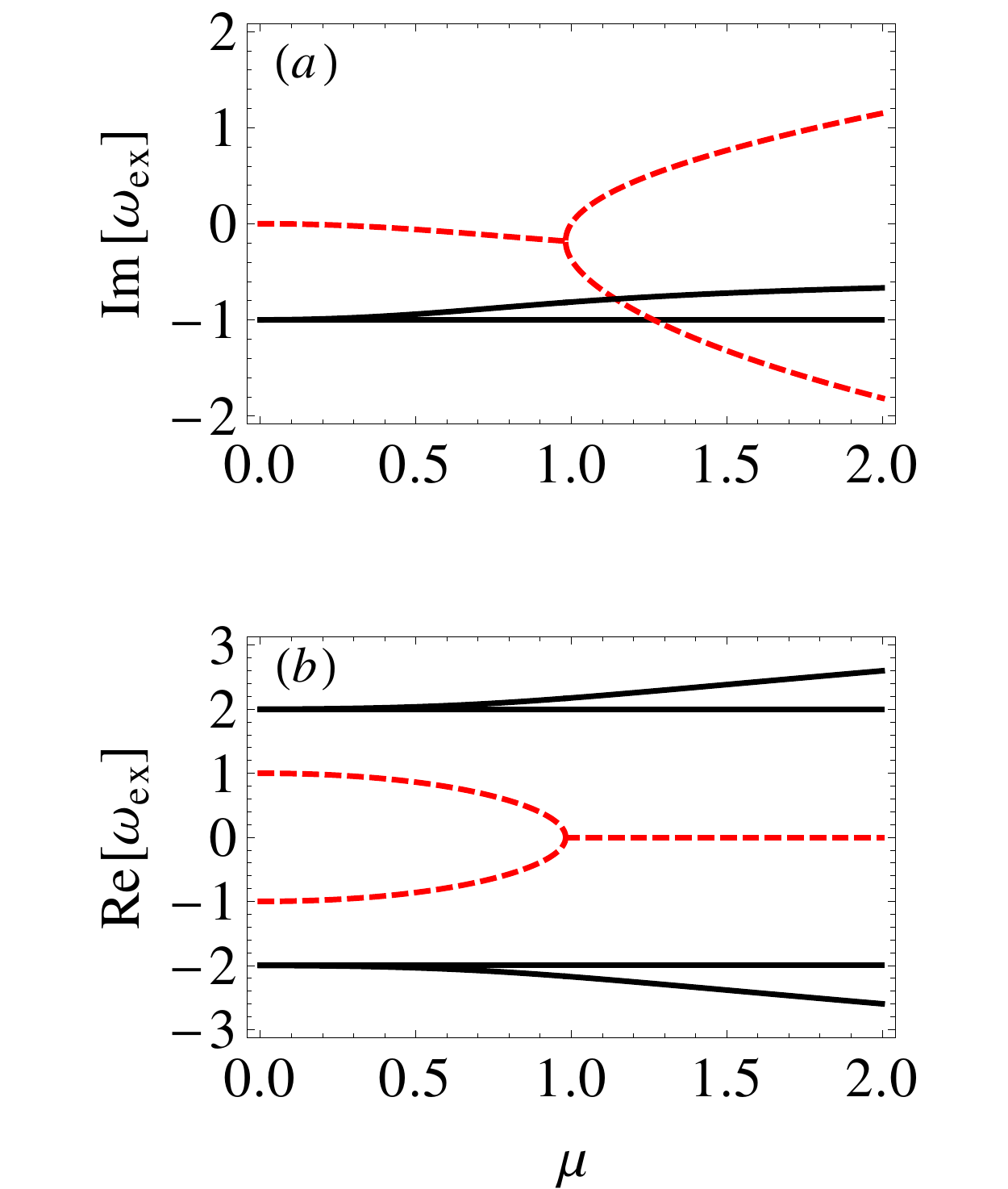}
\end{center}
\caption{Real and imaginary parts of the small excitation frequencies as a function of $\mu=
  \lambda/\lambda_{c}$ for $g = 2$, $\kappa =
  1$ and $\gamma=0$. In the limit of no light-membrane coupling, $\mu=0$, the solid black and dashed red curves correspond to the light and membrane modes, respectively.   We can see the
  effect of the critical coupling in the region of $\mu = 1$ where bifurcations
  take place for real and imaginary parts of the frequencies.}
	\label{fig:freqs}
\end{figure}

Let us switch to numerical analysis for the case $\kappa\neq0$.  Figure \ref{fig:freqs} plots the numerical results for the real and imaginary parts of the frequencies for the
steady states of the two optical fields and the membrane. The colour scheme used in the figure assigns colours depending  on the nature of the excitation  at $\mu=0$ where the radiation pressure on the membrane is switched off so that there is no backaction by light upon it. In this limit all the frequencies can therefore be assigned to pure excitations of either the light (solid, black) or the membrane (dashed, red). Despite the absence of backaction on the membrane at $\mu=0$, the membrane still acts on the light such that the global optical modes, which are the even and odd combinations of the left and right modes, are split by $2g$ which  is a quantity determined by the membrane's reflection and transmission coefficients.  Thus, at $\mu=0$ we see in panel (a) the decaying part of the optical mode at frequency $\kappa$, and in panel (b) we see that exciting a photon from the lower (antisymmetric) mode to the higher (symmetric) mode costs an energy to $2\hbar g$. 

We also see from Figure \ref{fig:freqs} that at $\mu=0$  the membrane
oscillates at its natural frequency without an imaginary part because we have excluded direct phonon loss, {\it i.e.} $\gamma=0$. However, as we increase $\mu$ modes become coupled and the membrane begins to pick up some damping from the light, which is the process responsible for membrane cooling \cite{jayich08}.  For $\mu > 1$ the motion of the membrane is no longer oscillatory and the fluctuations grow/decay
from/to the steady state.  We reiterate that Fig.\ \ref{fig:freqs} is
generated by expanding Eqns.\ (\ref{eq:lineom}) about the steady state
values of the three modes when $x = 0$. If we chose to expand around
one of the minima of the double well potential at $x \neq 0$ we would
find the opposite result, \textit{i.e.} small oscillations for $\mu > 1$
and exponential growth/decay for $\mu < 1$. We see that for non-zero $\kappa$ the initially degenerate light modes begin to split. One pair remains unchanged as we increase $\mu$ and continues to describe the single photon excitation process from the antisymmetric mode to the symmetric mode in the presence of a stationary membrane.  The other pair describes a similar process, but this time the membrane assists in the transition. Indeed, we can see that
in this case the decay rate of the ``light'' decreases slightly since some of its energy is now trapped in the interaction with the membrane and cannot leak out of the cavity and its
oscillations also increase the magnitude of their frequency.   Some of
the features mentioned are found in light-BEC coupled single cavity
systems used to model the Dicke PT.  Specifically, we see photon
assisted cooling of $N$ atoms in Figs.\ (5) and (1) in Refs.\
\cite{nagy2010,nagy11} and \cite{dickecrit0}, respectively.  We also see
increased oscillations of the photon mode as the coupling approaches
the critical point.  However, the oscillations in the single cavity
case correspond to the frequency of the light in the cavity and in our
case to tunnelling of photons between the two cavities. Finally, we emphasize that the results found in this section, especially the region around $\mu = 1$ shown in Fig.\ \ref{fig:freqs}, are best understood using the effective potential presented in Sec.~\ref{sec3}. Specifically, when the frequency of the primarily membrane mode goes to zero we get the formation of the double well in $V_{\mathrm{eff}}(x)$ where the membrane no longer oscillates about x = 0, but (assuming symmetry breaking) moves to one of the two new minima at $\pm x_{ss}$.

\section{Experimental Considerations and the Effects of Fluctuations}\label{sec6}

In this section we address the question of whether a Dicke type PT in a multimode optomechanical system is achievable with current experimental parameters. In addition, we shall also attempt an \emph{estimate} of the magnitude of the fluctuations of the light field which we have ignored in our mean-field treatment in order to check whether they are capable of averaging out the PT. In connection to this latter point, we note that fluctuations enter in two rather distinct ways. One way is as the critical fluctuations that occur when characteristic length and time scales diverge near the transition point even in a closed system. The other way is through the open nature of the cavity which leads to photon loss at random times. The former type of fluctuations are rather well understood and not particularly dangerous from our point of view as they are limited to the critical region and we have, {\it e.g.}, evaluated our critical exponents away from the critical point where mean-field theory is expected to be valid. We shall instead concentrate here upon the latter type which have the potential to invalidate a mean-field treatment even outside the critical region.

A simple estimate of photon number fluctuations can be found based
upon the observation, already made above in Section \ref{sec2}, that
the exact steady state solution of a driven and damped quantum
harmonic oscillator, that here represents the electromagnetic field
inside the cavity, is a coherent state \cite{louisell}. Even if the
coupling to the membrane changes the model from the pure harmonic
oscillator studied in \cite{louisell}, it should still be accurate to
assume a coherent state for the light field in the large field limit
we are interested in~\cite{irish}. We choose the amplitude of the
coherent state by matching it to the amplitude of our mean-field solution, in other words we assume that our mean-field solution for the intensity of the light field correctly gives the average number of photons in the cavity. Indeed, this connection is necessary if the quantum and classical approaches are to agree in the limit of large quantum numbers. This directly gives the size of the photon number fluctuations $\Delta N$ because for a coherent state we have  $\Delta N = \sqrt{N_{\mathrm{tot}}} =\sqrt{n_a + n_b}$.  These number fluctuations 
become dangerous in the broken symmetry state when they are of the same order of magnitude as the photon number difference between the left and right sides of the membrane, a quantity which acts as an order parameter.  We shall therefore define a ratio $\mathcal{R}$ analogous to a signal-to-noise ratio
\begin{equation}
\mathcal{R} = \frac{\left | n_a - n_b \right |}{\sqrt{n_a + n_b}} \, .
\label{eq:SNR}
\end{equation}  
We require $\mathcal{R} \gg 1$ for the open system fluctuations to have negligible effect. 
The mean-field expressions for the number difference and total number in the
symmetry broken phase are
\begin{eqnarray}
n_a - n_b &=& \pm \frac{V}{\lambda^2 \omega} \sqrt{2 \sqrt{g \omega}
  \eta \lambda - (g^2 + \kappa^2)} \\
n_a + n_b &=& \frac{\eta V}{\sqrt{g \omega} \lambda}
\end{eqnarray}  
where we have transformed back to dimensional quantities  $g \rightarrow g/\omega$,
$\kappa \rightarrow \kappa/\omega$ and $\eta \rightarrow \eta/\omega$,
in order to explicitly show the dependence upon the phonon frequency
$\omega$.  Combining these expressions with the expression for
$\lambda$ given in Eq.\ (\ref{eq:lambda}) gives
\begin{equation}
\mathcal{R} = \frac{\omega L}{\omega_{\mathrm{centre}}} \sqrt{\frac{g
    m V}{2 \hbar}} \sqrt{1 - \frac{1}{\mu_P}}
\end{equation}
where $\mu_P = \sqrt{\frac{P}{P_c}}$ and 
\begin{equation}
P_c = \frac{1}{16} \frac{\omega^2 L^2 m}{\omega_{\mathrm{centre}}}
\frac{(g^2 + \kappa^2)^2}{g \kappa} \ .
\end{equation}
In this last expression we have introduced the laser power $P$ and its critical value $P_{c}$. Although in the rest of this paper we have used $\lambda$ as the external parameter which is tuned across the critical point because it
makes the comparison to the DM most transparent, in an experiment $\lambda$ is not the most convenient parameter to tune as can be seen by examining the quantities entering its definition in Eq.\ (\ref{eq:lambda}). However, from
Eq.~(\ref{eq:crit}) we see that the critical coupling $\lambda_c$ is
inversely proportional to the pumping amplitude $\eta$ so that in an
actual experiment one would therefore probably choose $\eta$ as the control
variable.  Physically, we can control $\eta$ by tuning the power output of
the pumping laser(s), $P$, as we see by combining the relations $\eta =
\sqrt{\kappa I}$ and $P = I \hbar \omega_{\mathrm{centre}}$ where the photon current $I$
is the number of incident photons per second that match the cavity mode.

We shall now estimate $\mathcal{R}$ and $P_c$ using experimentally
realizable parameter values.  We first look at $g$ and note that in the
original membrane-in-the-middle experiment $g \sim 2
\pi \times 1$ GHz \cite{harris1}, but in more recent experiments it can be tuned to much smaller values $g \sim 2 \pi \times 0.1$ MHz \cite{sankey10}, giving a very large range of possibilities. The  membrane consists of a 50 nm thick film of SiN and has a fundamental mode of oscillation at a
frequency of $\omega \sim 2 \pi \times 100$ kHz \cite{zwickl08}. Its effective ``motional mass'' as appearing in Eq.\ (\ref{ham0}), which is one quarter of its actual mass for a square membrane, is quoted in reference \cite{harris1} as being $m=5 \times 10^{-14}$ kg and it sits in a cavity of length $L=0.067$ m.  Furthermore, high finesse
cavities have already been used to observe normal mode splitting, {\it
  i.e.} $\kappa < g$ \cite{sankey10} and in the future it is expected
that $\kappa \sim \omega$  \cite{marquardt13}. We set $\kappa \sim
\omega$ and assume that $\omega_{\mathrm{centre}}= c \times 2 \pi / 1064$ nm~\cite{harris1}.  Choosing $g=2\pi \times 10$ MHz and putting $V=1$ we arrive at the critical
values of $P_c \approx 1.2$ mW and  $\mathcal{R}\approx 625$ for $P = 1.1 \, P_c$.   These results for $\mathcal{R}$ and $P_{c}$ suggest both that fluctuations due to photon loss will not scramble the symmetry broken state and that the transition can be engineered to occur at reasonable parameter values, respectively. In terms of photon numbers inside the cavity the above parameter values  give $n_{a}+n_{b}= 2.2 \times 10^6$ and $n_{a}-n_{b}=9.3 \times 10^5$, and the number of phonons in the membrane is $n_{c}=1.0 \times 10^{7}$. The size of these numbers suggests that the mean-field approach should be valid at these parameter values. The value for $P_{c}$ we find is about ten times greater than the power used in the experiments \cite{harris1} and should be easily achievable. Our chosen value of $g$ sits in the middle of the experimentally realized range and increasing it to the upper end, \textit{i.e.} to $g \sim 2
\pi \times 1$ GHz,  would improve the situation for $R$ but would also increase $P_{c}$ by a factor of one million.

In the above calculation, and in this paper generally, we have ignored the mechanical damping of the membrane. The mechanical damping rate $\gamma$ of the fundamental mode depends on temperature
such that its quality factor $Q \sim 10^6$ at 293 K, and this increases to
$Q \sim 10^{7} $ at 300 mK \cite{zwickl08}. In units of the
membrane frequency we therefore have $\gamma = 1/Q  < 10^{-6}$. Multiplying by the above value of $n_{c}$ tells us that we lose about 10 phonons per natural period of the membrane in the symmetry broken state.  To put this in perspective we can consider  the power loss due to mechanical damping  $\gamma n_{c} \hbar \omega = 4.2 \times 10^{-22}$Js$^{-1}$ and compare it to that due to photon loss $\kappa (n_{a}+n_{b}) \hbar \omega_{\mathrm{centre}}=2.6 \times 10^{-7}$Js$^{-1}$. This shows that the phonon decay channel is energetically insignificant in comparison to the optical decay channel.

\section{Concluding remarks}\label{sec7}

In this work we have shown how dynamical critical phenomena may appear in the ``membrane in the middle'' optomechanical cavity setup. By first demonstrating that the corresponding closed model is identical to the DM we were led to analyze the driven-open system. At a mean-field level we showed that the model is indeed critical for certain drivings, meaning that the steady state solutions of the classical equations of motion exhibit a bifurcation which signals a second order PT. We calculated two critical exponents: $\alpha$ which characterizes the critical behaviour of the ground state wave function of the membrane as the light-membrane coupling strength $\lambda$ is varied, and $\beta$ which is the related quantity for the number of membrane phonons, and found they were identical to those of the DM. To be more precise,  we found that $\alpha$ was identical to that of the closed DM (to the best of our knowledge it has not yet been calculated for the open DM, but since it does not depend on $\kappa$ we do not expect any difference for the open and closed DM's anyway), and we found that $\beta$, which does differ between the open and closed DM's, agrees with that of the open DM. The structure of the spectrum of quantum fluctuations about the mean-field also looks identical to that of the DM suggesting a connection also at the quantum level. Inputting experimentally realistic numbers showed both that the PT was within reach with current technology and that the mean-field theory can be a reasonable approximation, at least outside of the usual critical region.

The realization of a Dicke-type PT in a macroscopic mechanical system would be a significant step.
The experiments that have seen the Dicke PT so far have all been with atomic gases: the dynamical Dicke PT has been demonstrated in cold atom systems where the atoms spontaneously choose one of two spatial configurations~\cite{black03,beccav2}, and very recently the PT has been seen in a thermal gas of $\Lambda$-atoms Raman coupled to an external drive and a cavity mode~\cite{dickerel2}, and also in a spin-orbit coupled atomic condensate~\cite{dickeso}. As our results give strong evidence that the multimode optomechanics model belongs to the same universality class as the DM, they open up a new arena where critical Dicke physics could be explored.  Of course, the mean-field treatment in this paper is only the first step in this direction, and future theoretical work should probably attempt to verify the coherent state hypothesis in Sec.\ \ref{sec6} which estimates the effect of fluctuations due to the open nature of the system and indicates they should not be an obstacle to observation of a PT. 

In closing, we note that the realization of a PT in the atomic gas experiments \cite{black03,beccav2,beccav3} cleverly overcame two serious obstacles. The first is that at optical frequencies the critical coupling strength $\lambda_{c}$  obtained from the closed undriven Dicke Hamiltonian Eq.\ (\ref{dicke}) becomes equal to the optical frequency (when units are restored). This is unfeasibly large (much larger than the dipole coupling), seemingly making the superradiant state unreachable. The second is the existence of a {\it no-go theorem}~\cite{rzazewski75,bialynicki1979no} which states that the Dicke PT is not allowed as an equilibrium PT for two-level atoms coupled to a single photon mode. The origin of this no-go theorem can be understood from the minimal coupling Hamiltonian $(\vec{p}-e\vec{A})^2$ that gives the coupling between the atoms and the light: the original DM had the term $\vec{p} \cdot \vec{A}$ but left out the field self-energy $\propto A^{2}$ which when included prevents the PT. The solution to both these problems is to use an open system which employs a Raman scheme involving an external pump laser (as well as the cavity field) \cite{dimer07}. This not only breaks the relation between the coupling strength and the magnitude of the self-energy implied by the minimal coupling Hamiltonian, but also allows the effective two-level system coupled by the Raman scheme to be formed from two states in the ground state manifold with an arbitrarily small splitting meaning that the optical frequency is no longer the relevant frequency scale. In an analogous fashion to the atomic systems, the possibility of a PT in our model follows from the fact that the smallness of the tunneling rate $g$ in comparison to the membrane frequency $\omega$ is in principle not limited by some physical constraint, allowing us to tune the parameters to achieve a critical coupling $\lambda_{c}$ which is readily achievable in experiments. The no-go theorem is likewise circumvented.

\begin{acknowledgements}
JM and DHJO'D thank NSERC (Canada), and JL thanks VR-Vetenskapsr\aa det for financial support.
\end{acknowledgements}

\appendix

\section{Sensitivity of Schr\"{o}dinger Cat States to Asymmetry}
\label{sec:appendix}

In Section \ref{sec3} we discussed the possibility of a macroscopic superposition (Schr\"{o}dinger Cat state) of two membrane positions centred at the minima of the effective double well potential. In this appendix we shall pick one possible decoherence mechanism, namely an asymmetry in the double well potential, and determine the sensitivity of such a state to it.  The asymmetry might arise from an imbalance in the pumping, or it might arise from other effects we have not explicitly considered in which case imbalanced pumping might even provide a way to try and counter it. What is guaranteed, however, is that for a large enough separation of the minima any infinitesimal imbalance will cause the collapse of the state into one well or the other.  Since the imbalanced pumping appears as the odd term in $V_{\mathrm{eff}}$ in Eq.\ (\ref{eq:Veff}) it provides a precise way to examine the effects of such imbalances whatever their physical origin.  

Above the PT and for $\eta_a = -\eta_b$ we have a perfectly balanced double well potential. In the absence of tunnel coupling between the two wells the ground state would be doubly degenerate with energy $\mathcal{E}$, say, and the two ground state wave functions $\psi_L(x)$ and $\psi_R(x)$ are localized in the left and right wells, respectively. States localized in the left or right well are of course closest to the classical description. In the presence of tunnelling the degeneracy is lifted and to a very good approximation the new ground and first excited state are given by  (see Problem 3 on p 183 of reference \cite{landau&lifshitz}) 
\begin{eqnarray}
E_g: \hspace{5mm} \psi_g(x) &=& \left [\psi_L(x) + \psi_R(x) \right
]/\sqrt{2} \, , \nonumber \\
E_e: \hspace{5mm} \psi_e(x) &=& \left [\psi_L(x) - \psi_R(x) \right
]/\sqrt{2} \, .
\end{eqnarray}
The splitting $\Delta E = E_{e} - E_{g}$  determines the sensitivity of the system to small imbalances in the double well potential. When the tilt in $V_{\mathrm{eff}}$ is much smaller than $\Delta E$, the symmetric state remains the ground state. When the tilt is of order of or larger than $\Delta E$ the ground state becomes localized in the lower well.
To estimate the splitting one can use the WKB result which, when translated into our units, reads
\begin{equation}
\Delta E = \frac{\Omega}{\pi} \exp \left[-\sqrt{2}
  \int_{-a}^{a} \sqrt{V_{\mathrm{eff}}(x) - \mathcal{E}} \, dx \right]
\label{WKB}
\end{equation}
where $\pm a$ are the positions of the classical turning points on either side of the central barrier and $\Omega$ is the angular frequency (in units of $\omega$) of the classical motion having energy $\mathcal{E}$ in either well.  In the harmonic approximation, we find that the potential near the bottom of the wells is given by
\begin{eqnarray}
V_{\mathrm{eff}} & \approx & 2 \frac{\mu-1}{\mu}(x-x_{ss})^2+E_{0} \\
 & \equiv & \frac{1}{2} \Omega^{2}(x-x_{ss})^2  +E_{0}
\end{eqnarray} 
where $E_{0}$ is the energy at the bottom of the well given in Eq.\ (\ref{eq:GSEnergy}). This implies that the quantum ground state energy in each well is given by
\begin{equation}
\mathcal{E} \approx \sqrt{\frac{\mu-1}{\mu}}+E_{0}.
\end{equation}
Next, we approximate the tunnel barrier by an inverted parabola and find that near $x=0$
\begin{equation}
V_{\mathrm{eff}} \approx \epsilon_{0} -\frac{1}{2}(\mu^{2}-1)x^{2}
\end{equation}
where $\epsilon_{0}$ is defined below Eq.\ (\ref{eq:x}).
The turning points $\pm a$ of the classical motion at the central barrier can now be estimated by equating this potential to $\mathcal{E}$ giving
\begin{equation}
a^{2}=\frac{1}{1+\mu} \left[ 2 \epsilon_{0}\frac{\mu-1}{\mu^2}- 2 \sqrt{\frac{1}{\mu (\mu-1)}} \right] . 
\end{equation}
Combining these results we arrive at the expression
\begin{equation}
\Delta E \approx   \frac{2}{\pi}\sqrt{\frac{\mu-1}{\mu}} \exp\left[ \frac{-\pi \left( \epsilon_0
      \frac{(\mu - 1)^2}{\mu^2} - \sqrt{\frac{\mu - 1}{\mu}}  \right)}{\sqrt{\mu^2 - 1}}\right] .
\end{equation}

 Due to the exponential smallness of the tunnel splitting even a tiny imbalance in the double well will make a big difference. The energy imbalance associated with unequal pumping is given by the  third term in Eq.\ (\ref{eq:Veff}) for $V_{\mathrm{eff}}$.   To keep the calculation as simple as possible we expand this term to first order in $x$ and evaluate it at the minimum energy of the unperturbed double well, 
 \begin{equation}
\Delta E_{\mathrm{Imb}} = 2 V (\eta_a^2 - \eta_b^2) \frac{(g^2 -
  \kappa^2) \sqrt{\mu - 1}}{(g^2 + \kappa^2)^{3/2}}  \, . 
\end{equation}
We have included a factor of $2$ to take account of the difference in energy between the two wells.
Equating this to $\Delta E$ gives the relation
\begin{equation}
\eta_a^2 - \eta_b^2 = \frac{(g^2 + \kappa^2)^{3/2}}{V \pi  \sqrt{\mu} (g^2 -
  \kappa^2)} \, \exp\left[ \frac{-\pi \left( \epsilon_0
      \frac{(\mu - 1)^2}{\mu^2} - \sqrt{\frac{\mu - 1}{\mu}}  \right)}{\sqrt{\mu^2 - 1}}\right] .
\end{equation}
Switching to dimensional quantities and expressing in terms of the power difference (see Section \ref{sec6}) gives
\begin{equation}
\Delta P = P_a - P_b = \mathrm{C}_1 \, \mathrm{exp} \left[ -\pi \frac{\left( \frac{\mathrm{C_2}}{\hbar \omega}
      (\mu_P - 1)^2 - \sqrt{\frac{\mu_P - 1}{\mu_P}}  \right)}{\sqrt{\mu_P^2 - 1}} \right]
\label{powerdiff}
\end{equation}
where 
\begin{eqnarray}
\mathrm{C}_1 &=& \frac{\hbar \omega \omega_{\mathrm{centre}}
  (g^2 + \kappa^2)^{3/2}}{V \pi \kappa \sqrt{\mu_P} (g^2 - \kappa^2)} \, , \nonumber \\
\mathrm{C}_2 &=& \frac{V L^2 \omega^2 m (g^2 + \kappa^2)}{8
  \omega_{\mathrm{centre}}^2} \nonumber \, .
\end{eqnarray}
Inputting the experimental parameters we used in Sec.\ \ref{sec6} into the RHS
of Eq.\ (\ref{powerdiff}) gives a splitting of $\Delta P \propto 10^{-2.1 \times 10^{6}}$ which definitively excludes the survival of a cat state.  Very near the PT the central barrier in the double well potential gets smaller and the system is less sensitive to an imbalance. For example, at $P = 1.00001 \, P_c$ we find
$\Delta P \approx 0.2$pW, which is more viable than the previous case but then the cat state must also contend with
fluctuations in the critical region of the PT. We therefore conclude that the observation of a cat state in current experiments is virtually impossible  but future experiments with lighter and more reflective membranes interacting with high frequency light in smaller cavities with lower loss rates might ease the situation.

\end{document}